\documentclass[preprint2]{aastex}

\usepackage{graphicx}
\usepackage{graphics}
\usepackage{txfonts}
\usepackage{epsfig}
\usepackage{natbib}
\usepackage{times}
\usepackage{amssymb}
\usepackage{color}
\usepackage{longtable}

\newcommand\phcms{ph\,cm$^{-2}$\,s$^{-1}$}

\newcommand\cps{cts\,s$^{-1}$}
\newcommand\cmsq{cm$^{-2}$}

\newcommand\fermi{{\it{Fermi}}}
\newcommand\agile{{\it{AGILE}}}
\newcommand\rxte{{\it{RXTE}}}
\newcommand\maxi{{\it{MAXI}}}
\newcommand\swift{{\it{Swift}}}
\newcommand\cygi{{\object{Cyg~X-1}}}
\newcommand\cygiii{{\object{Cyg~X-3}}}
\newcommand\grs{{\object{GRS~1915$+$105}}}
\newcommand\gx{{\object{GX~339$-$4}}}

\newcommand\psr{{\object{PSR~J2032$+$4127}}}
\newcommand\lsi{{\object{LS~I$+$61$^{\circ}$303}}}
\newcommand\lsii{{\object{LS~5039}}}

\shorttitle{\emph{Fermi}-LAT observations of microquasars}
\shortauthors{Bodaghee et al.}

\begin{document}

\title{Gamma-ray observations of the microquasars Cygnus~X-1, Cygnus~X-3, GRS~1915+105, and GX~339$-$4 with the \emph{Fermi} Large Area Telescope}

\author{Arash Bodaghee}
\affil{\footnotesize{Space Sciences Laboratory, 7 Gauss Way, University of California, Berkeley, CA 94720, USA}}
\email{\footnotesize{bodaghee@ssl.berkeley.edu}}


\author{John A. Tomsick}
\affil{\footnotesize{Space Sciences Laboratory, 7 Gauss Way, University of California, Berkeley, CA 94720, USA}}


\author{Katja Pottschmidt}
\affil{\footnotesize{CRESST and NASA Goddard Space Flight Center, Astrophysics Science
Division, Code 661, Greenbelt, MD 20771, USA \\
Center for Space Science and Technology, University of Maryland Baltimore
County, 1000 Hilltop Circle, Baltimore, MD 21250, USA}}


\author{J\'{e}r\^{o}me Rodriguez}
\affil{\footnotesize{Laboratoire AIM, CEA/IRFU - Universit\'e Paris Diderot - CNRS/INSU, \\ CEA DSM/IRFU/SAp, Centre de Saclay, F-91191 Gif-sur-Yvette, France}}


\author{J\"{o}rn Wilms }
\affil{\footnotesize{Dr. Karl Remeis-Sternwarte and Erlangen Centre for Astroparticle Physics, Friedrich-Alexander-Universit\"{a}t Erlangen-N\"{u}rnberg, Sternwartstra$\beta$e 7, 96049 Bamberg, Germany}}


\author{Guy G. Pooley}
\affil{\footnotesize{The University of Cambridge, Mullard Radio Astronomy Observatory, Cavendish Laboratory, \\J.J. Thomson Avenue, Cambridge CB3 0HE, UK}}

\begin{abstract}
Detecting gamma-rays from microquasars is a challenging but worthwhile endeavor for understanding particle acceleration, the jet mechanism, and for constraining leptonic/hadronic emission models. We present results from a likelihood analysis on timescales of 1\,d and 10\,d of $\sim$4 years worth of gamma-ray observations (0.1--10 GeV) by \emph{Fermi}-LAT of Cyg X-1, Cyg X-3, GRS 1915$+$105, and GX 339$-$4. Our analysis reproduced all but one of the previous gamma-ray outbursts of Cyg X-3 as reported with \emph{Fermi} or \emph{AGILE}, plus 5 new days on which \cygiii\ is detected at a significance of $\sim$5$\sigma$ that are not reported in the literature. In addition, \cygiii\ is significantly detected on 10-d timescales outside of known gamma-ray flaring epochs which suggests that persistent gamma-ray emission from \cygiii\ has been detected for the first time. For \cygi\, we find three low significance excesses ($\sim$3--4$\sigma$) on daily timescales that are contemporaneous with gamma-ray flares reported (also at low significance) by \emph{AGILE}. Two other microquasars, GRS 1915+105 and GX 339$-$4, are not detected and we derive 3$\sigma$ upper limits of $2.3\times10^{-8}$\,\phcms\ and $1.6\times10^{-8}$\,\phcms, respectively, on the persistent flux in the 0.1--10\,GeV range. These results enable us to define a list of the general conditions that are necessary for the detection of gamma-rays from microquasars. 
\end{abstract}


\section{Introduction}

A microquasar ($\mu$QSO) consists of a compact object (CO: a neutron star, NS, or a black hole, BH) that accretes matter from a normal stellar companion. The characteristic that distinguishes $\mu$QSOs from other X-ray binaries (XRBs) is the presence of non-thermal synchrotron emission from relativistic jets launched near the CO \citep{mir99}. These radio jets are believed to be powered by black-hole spin and/or by strong electromagnetic currents in the inner accretion disk \citep[e.g.,][]{bla77,mei01}. 

The presence of jets interacting with plasmas within strong electromagnetic and gravitational fields leads $\mu$QSOs to display rapid variability across a broad range of frequencies: from radio, to IR, and X-rays. A few of these objects have been detected in the gamma-rays ($>$100\,MeV) with \emph{AGILE} \citep{tav09a}, \emph{Fermi}-LAT \citep{atw09}, and MAGIC \citep[][and references therein]{lor04}. These are \cygiii\ \citep{corbel09,tav09b,corbel12,pia12} and \cygi\ \citep{alb07,bul10a,sab10a,sab10b,sab13}. While the XRBs \object{LS~I$+$61$^{\circ}$303} \citep{alb06,2009ApJ...701L.123A} and \object{LS~5039} \citep{par04,aha05a,2009ApJ...706L..56A} have been detected at MeV--TeV energies, it is still uncertain whether they should be classified as $\mu$QSOs \citep[e.g.,][]{par11a}.

Leptonic and hadronic processes are generally invoked to explain gamma-ray emission from $\mu$QSOs. In the former, relativistic electrons in the jet emit synchrotron radiation (with some loss due to self-absorption), or they can Compton upscatter low-energy (IR and UV) photons from the accretion disk or from the stellar companion to gamma-ray energies \citep[e.g.,][and references therein]{kau02,rom02,bos06,sit12}. In hadronic models, inelastic collisions between jet protons and those of the dense stellar wind produce neutral pions which decay into gamma-rays and neutrinos \citep[e.g.,][and references therein]{rom03}. Interactions between the jet and the clumpy winds from massive donor stars in HMXB microquasars can lead to gamma-ray emission from both leptonic (IC) and hadronic (neutral pion decay) processes \citep{ara09,owo09}. These  winds can also serve as the site for the initiation of e$^{-}$-e$^{+}$ cascades from a primary source of very-high energy (VHE) gamma-rays within the system; the secondary emission resulting from these pair cascades should be detectable in the low-energy gamma-rays \citep{bed97,rom10}. Some of these photons can be absorbed by the wind leading to variable gamma-ray emission \citep{dub06}. Shocks at the termination zone where the jet meets the interstellar medium are also believed to produce gamma-ray photons \citep{bos11}.

\begin{deluxetable}{ l c c l c c }
\tablewidth{0pt}
\tablecaption{Microquasar candidates selected for this study.}
\tablehead{
\colhead{Name} & \colhead{R.A. (J2000) [deg]}  & \colhead{Decl. (J2000) [deg]}  & \colhead{$P_{\mathrm{orb}}$\,[d]} & \colhead{Distance\,[kpc]}	& \colhead{Type \tablenotemark{\dagger}} }
\startdata

\object{Cygnus~X-1} 		& 299.590 			& $+$35.202 			& 5.599829(16)$^{a}$	& $1.86_{-0.11}^{+0.12}$$^{b}$	& HMXB	\\
\object{Cygnus~X-3} 		& 308.107 			& $+$40.958 			& 0.1996907(7)$^{c}$	& 7.2(5)$^{d}$			& HMXB	\\
\object{GRS~1915$+$105} 	& 288.798 			& $+$10.946			& 33.5(1.5)$^{e}$		& $11_{-4}^{+1}$$^{f}$	& LMXB	\\
\object{GX~339$-$4} 		& 255.706				& $-$48.790			& 1.7563(3)$^{g}$		& $10_{-4}^{+5}$$^{g}$	& LMXB	\\

\enddata
\tablenotetext{\dagger}{HMXB: high-mass X-ray binary; LMXB: low-mass X-ray binary}

\tablerefs{
($a$) \citet{bro99}; ($b$) \citet{rei11}; ($c$) \citet{wen06}; ($d$) \citet{lin09}; ($e$) \citet{har04}; ($f$) \citet{jon04}; ($g$) \citet{hyn04}.}

\label{tab_targets}
\end{deluxetable}

The common trait shared by these emission models is that gamma-rays are tied to the presence of radio jets \citep{par11b}. These jets appear during specific emission states for black-hole XRBs \citep[BHXBs: e.g.,][and references therein]{fen04,mcc06}. Indeed, this link between radio and gamma-ray emission was demonstrated for several gamma-ray outbursts of \cygiii\ \citep{corbel10,wil11,corbel12}. It is clear then that observations of this class of objects in the gamma-rays can shape our understanding of the role jets play in particle acceleration and the production of high-energy photons.

In this paper, we present a systematic study of four well-known $\mu$QSOs (\cygi, \cygiii, \grs, and \gx) using gamma-ray data collected by the Large Area Telescope on board the \emph{Fermi} space telescope. In \S\,\ref{sec_obs}, the sources selected for this study are introduced, and the dataset and analysis techniques are described. We present results for individual objects in \S\,\ref{sec_res}, and we discuss their implications in \S\,\ref{sec_disc}. A summary of our key findings is given in \S\,\ref{sec_conc}.

\section{Source List \& Data Analysis}
\label{sec_obs}

\subsection{Source List}

Our source list consists of the four $\mu$QSOs presented in Table\,\ref{tab_targets}. All four systems happen to host black hole candidates even though $\mu$QSOs can sometimes host neutron stars \citep[][and references therein]{fen02}. In recent years, these $\mu$QSOs have displayed a wide range of X-ray luminosities and state changes, and so they are the regular subjects of observations by telescopes in all wavelengths. Of these sources, only Cygnus~X-3 was unambiguously detected in the gamma-rays (with \agile\ and \fermi). There are reports of possible gamma-ray emission from \cygi\ detected by \agile\ \citep{bul10a,sab10a,sab10b,sab13} but the detection significances are low ($\sim$3--4$\sigma$) and these have not yet been corroborated with \fermi\ detections. The primary objectives of this work are to reproduce previous gamma-ray detections of \cygiii\ with the \fermi-LAT, and to apply a systematic search for possible gamma-ray emission from other $\mu$QSOs. A brief description of these $\mu$QSOs is provided in the following paragraphs.

\subsubsection{Cygnus~X-3}

Discovered in the X-rays by \citet{gia67}, \cygiii\ was subsequently detected in the radio band \citep{bra72}, and has since been observed and extensively studied in all wavelengths, including in the gamma-rays \citep[e.g.]{corbel09,tav09b,wil11}. The compact object in \cygiii\ is believed to be a black hole \citep[e.g.][]{sch10,zdz13} in a tight orbit around a Wolf-Rayet star \citep{van92}. This configuration (a relativistic jet near a dense stellar wind) leads to gamma-ray emission that is modulated with the 4.8\,h orbital period \citep{corbel09,dub10}. 

Around a dozen outbursts from this source have been detected above 100\,MeV by \fermi\ \citep{corbel12} and \agile\ \citep{pia12}. Multi-wavelength analysis of these gamma-ray outbursts led \citet{corbel12} to propose a set of three conditions that appear to be necessary for the detection of gamma-rays with the LAT: the source count rate in the \rxte-ASM 3--5\,keV band should be higher than about 3\,\cps; the count rate in the \swift-BAT 15--50\,keV band must cross (in either direction) the critical value of 0.02\,\cps; and there should be significant emission in the radio (flux densities $\gtrsim$0.2--0.4\,Jy). Whenever gamma-rays from this source are detected by the LAT, it is contemporaneous, but not necessarily coincident, with these conditions being satisfied. The order in which these various components (radio, X-rays, and gamma-rays) are emitted is not well understood with a time lag of $\Delta t = 5\pm7$\,d between the peak emission in the radio band and in the gamma-rays \citep{corbel09}.

\subsubsection{Cygnus~X-1}

Located at a parallax-derived distance of $1.86_{-0.11}^{+0.12}$\,kpc \citep{rei11}, \cygi\ \citep{bow65} is a bright, nearby $\mu$QSO featuring a likely black hole accreting from a high-mass companion with a binary orbital period of 5.6\,d \citep[e.g.,][]{gie08}. \cygi\ is a persistent X-ray emitter: its X-ray evolution does not follow a q-shaped track in the hardness-intensity diagram as is typically seen in transient systems \citep[e.g.,][]{fen04,wil06}. Instead, the source shows very fast (or sometimes failed) state transitions \citep[e.g.,][]{boc11,gri13}. This object is one of the best-studied X-ray sources in the Galaxy with firm detections from the radio band to the X-rays, and in the soft gamma-rays ($\lesssim$10\,MeV) with \emph{COMPTEL} \citep{mcc02} and with \agile\ \citep[$>$30\,MeV:][]{bul10a,sab10a,sab10b,sab13}. Polarized gamma-ray emission up to 2\,MeV, whose likely origin is the jet, was detected with \emph{INTEGRAL} \citep{lau11}. At even higher energies ($>$100\,GeV), a brief 4$\sigma$ excess was reported with MAGIC \citep{alb07} coincident with the beginning of a long-term brightening in the X-rays.

\subsubsection{GRS~1915+105}

An accreting black hole candidate in an LMXB system with a 33-d orbital period \citep[e.g.,][]{har04}, \grs\ has been in outburst since 1992 when it was discovered in the hard X-rays by \emph{Granat} \citep{cas92}. Apparent superluminal motion in the ejecta of its radio jets is one remarkable feature of this object \citep{mir94}. Another interesting feature is its wide range of X-ray variability classes \citep[e.g.,][]{bel00,kle02,han05}. Thus, \grs\ is one of the brightest and most variable sources in the X-ray sky making it a prime target for multi-wavelength campaigns from the radio to the hard X-ray band seeking to understand accretion physics around black holes and the interactions between the accretion disk and the relativistic jets \citep{rod08a,rod08b}. Non-thermal emission (i.e., from the jets or at the termination shock) could extend up to very high-energy (VHE) gamma-rays, however, such emission has not been detected from this source with, e.g., HESS \citep{szo09}, nor has it been detected in soft gamma-rays with \fermi\ or \agile.

\subsubsection{GX~339$-$4}

A recurrent X-ray transient, \gx\ \citep{mar73} is composed of a black hole candidate in a 1.8-d orbit \citep{jon04} around a low-mass donor star \citep{hyn03,hyn04}. The source undergoes regular outbursts in the X-rays which trace a $q$-track in the hardness-intensity diagram \citep{bel05}. This system features powerful relativistic jets that are suppressed in the high-soft state \citep{corbel00} leading to the discovery of a strong correlation between radio and hard X-ray emission \citep{fen99,corbel03}. The distance to \gx\ is not well known \citep[see discussion in, e.g.,][]{zdz04} but is suspected of being between 6\,kpc and 15\,kpc \citep{hyn04,zdz04}. Gamma-ray emission is expected from this $\mu$QSO \citep[e.g., see discussion in][]{vil10} but efforts to detect this emission have been unsuccessful.

\begin{figure*}[!t] \centering
\includegraphics[width=16cm,angle=0]{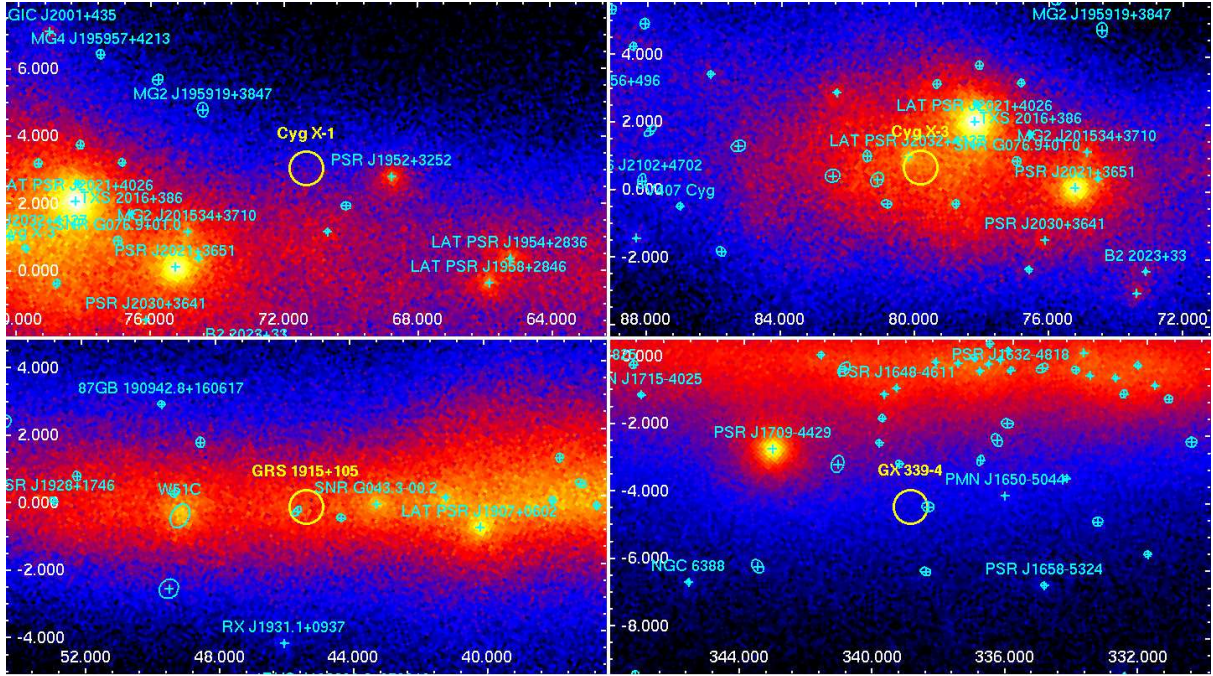}
\caption{Photon counts map (in Galactic coordinates) of an $18^{\circ}\times10^{\circ}$ field around the microquasars in our study from LAT data (0.1--10\,GeV) spanning 2008 August 5 through 2012 May 15. Galactic coordinates are shown as are the locations of sources from the 2FGL catalog \citep{lat11} that are included in the likelihood analysis; the symbol size represents the 95-\% error ellipse. The position of each microquasar is indicated by a $1^{\circ}$-diameter circle.}
\label{fig_cmap}
\end{figure*}

\begin{figure*}[!t] \centering
\includegraphics[width=16cm,angle=0]{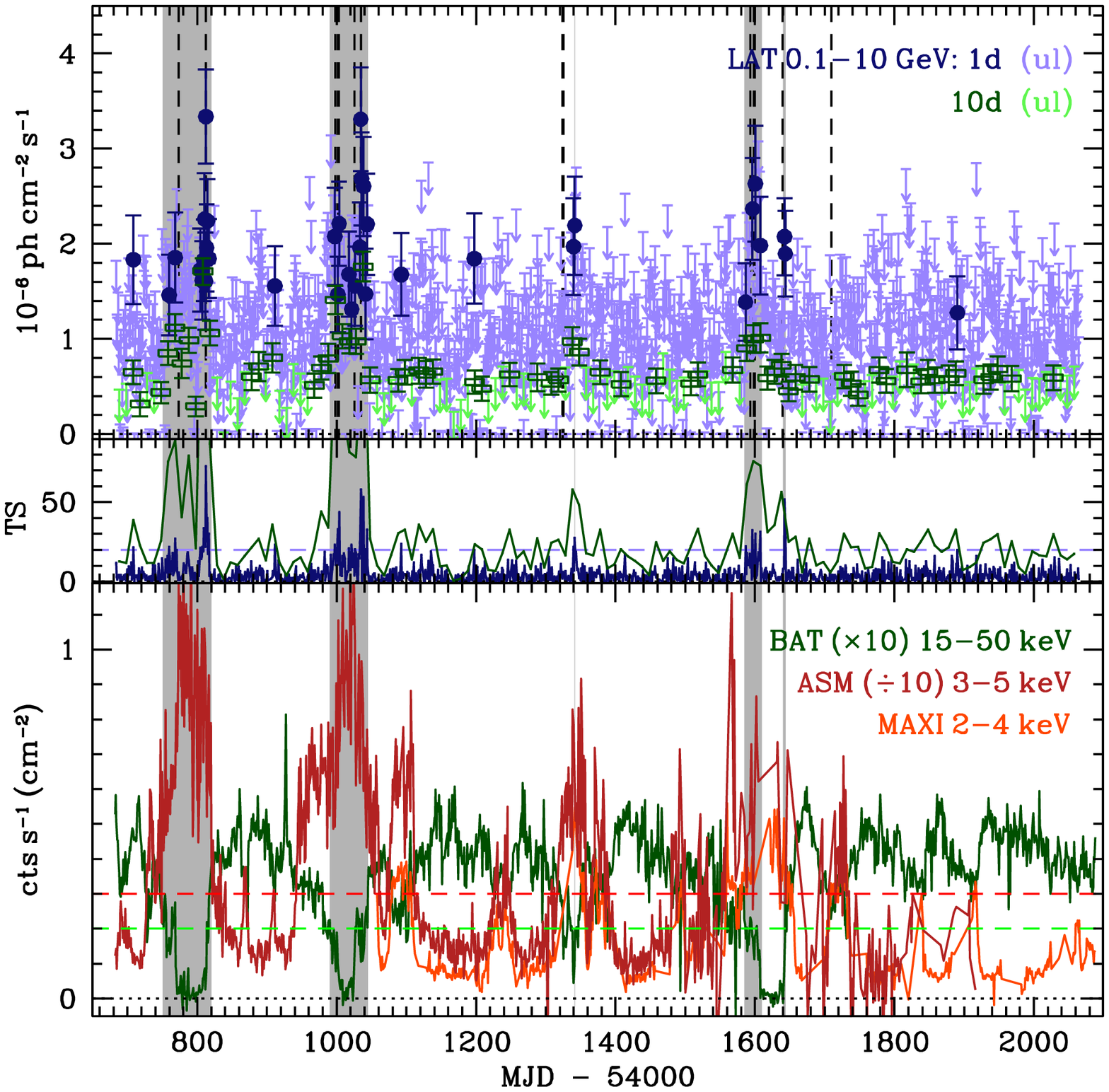}
\caption{Light curve of Cyg~X-3 as collected from a likelihood analysis of \fermi-LAT (0.1--10\,GeV) data taken between 2008 August 5 and 2012 May 15. Results are presented for 1-day and 10-day bins where the top panel presents the integrated flux (in $10^{-6}$\,\phcms) while the middle panel gives the corresponding test statistic (TS). Bins in which TS$\ge$20 (horizontal dashed line in the middle panel) are represented by disks (1-day) and rectangles (10-day) in the top panel, while downward arrows represent 1-$\sigma$ upper limits for bins with TS$<$20. Previously-reported gamma-ray detections by \agile\ (vertical dashed lines in the top panel) and by \fermi-LAT (shaded regions) are also shown. The bottom panel presents the daily light curves in soft and hard X-rays from \emph{MAXI}-GSC (\cps\,\cmsq\ in 2--4\,keV), \emph{RXTE}-ASM (\cps\ in 3--5\,keV), and \swift-BAT (\cps\,\cmsq\ in 15--50\,keV). In this panel, the horizontal dashed lines denote the X-ray thresholds defined by \citet{corbel12} for its detection in gamma-rays, i.e., an ASM count rate $\geq$3\,\cps, and a BAT count rate crossing 0.02\,\cps, respectively. Error bars have been omitted for clarity.} 
\label{fig_lc_CygX3}
\end{figure*}

\subsection{Data Analysis}

\fermi\ Science Tools v9r27 and HEASoft 6.12 were used to analyze all photon events within 20$^{\circ}$ of the sources listed in Table\,\ref{tab_targets} that were captured by the \fermi\ Large Area Telescope \citep[LAT:][]{atw09} between the first full day of data availability (2008 August 5: MJD\,54683) and 2012 May 15 (MJD\,56062). Good time intervals were selected for two energy bands (100\,MeV--1\,GeV and 100\,MeV--10\,GeV) with with \texttt{gtselect} and \texttt{gtmktime}. The events class was set to ``2'' which selects only high-quality diffuse photons. To minimize background albedo photons from the Earth's limb, we restricted the zenith and rocking angles to less than 100$^{\circ}$ and 52$^{\circ}$, respectively. The scripts \texttt{gtexpcube} and \texttt{gtexpmap} generated exposure maps while \texttt{gtbin} created photon counts maps in the region of interest. The instrument response functions consisted of \texttt{P7SOURCE\_V6}. 

To create light curves on short timescales, the spatial distribution of known sources and their spectral parameters must be modeled as accurately as possible. Counts maps created from the full data set are presented in Fig.\,\ref{fig_cmap}. Visual inspection of these fields reveals the challenge in assigning photons to our targeted sources. 

The BHXB state with the highest X-ray luminosity is usually the steep-power-law state \citep{mcc06} which features a photon index $\Gamma>2.4$, and a variable hard tail that can extend to $\sim$1 MeV \citep{mcc06}. While the X-rays and gamma-rays could have different power-law slopes \citep{corbel12}, we note that a photon index $\Gamma=2.7$ was measured with \fermi-LAT during gamma-ray outbursts of \cygiii\ \citep{corbel09}. This suggests that our targets should be faint, and any emitted gamma-rays are more likely to be found at the low end of the \fermi\ energy band ($\lesssim$1\,GeV). At these energies, the LAT point spread function (PSF) is of the order of a degree or more. Photons emanating from our targets (and from other sources) are spread across the images (and partially beyond). Our targets are situated near the Galactic Plane where there is a high level of diffuse background emission whose intensity peaks in the energy range of interest. Their spatial and spectral distributions are not well understood but they must be accounted for, nonetheless. 

For each target, we began with a model file that included all 2FGL sources \citep{lat11} located up to 5$^{\circ}$ away from the target position, and all bright (detection significance $> 7\sigma$ and flux ($>$100\,MeV) $> 5\times10^{-8}$\,\phcms) 2FGL sources up to 20$^{\circ}$ away. We also included the latest spectral and spatial models for the Galactic (\texttt{gal\_2yearp7v6\_v0}) and extragalactic\linebreak(\texttt{iso\_p7v6source}) diffuse emission. A binned likelihood analysis (\texttt{gtlike}) was performed on the full data set (0.1--1\,GeV and 0.1--10\,GeV) where sources within 3$^{\circ}$ of the target (including the diffuse emission components) had spectral parameters that were allowed to vary, while sources within 9$^{\circ}$ of our target had free normalizations. This enabled the spectral parameters of sources in the field to be constrained. A new model file was made in which all spectral parameters were fixed to the values derived from this binned analysis, except for the photon index and normalization of the target $\mu$QSO which were left free to vary. 

Then, we performed an unbinned likelihood analysis in the 0.1--1 and 0.1--10\,GeV energy range in three time bins: 0.1, 1, and 10 days. Only photons inside a circle of 7$^{\circ}$ radius centered on the target were considered in the likelihood analysis. Restricting the energy range to lower values (0.1--1\,GeV) increases the probability that the gamma-ray photons originate from our target, since these sources are expected to emit few photons above 1\,GeV. However, any gain in likelihood is offset by a higher background due to the larger size of the LAT PSF at low energies. There are no significant differences in the light curves generated from the likelihood analysis in the 0.1--1\,GeV and 0.1--10\,GeV range (other than the aforementioned higher level of background in the former), so only the 0.1--10\,GeV results are discussed hereafter. We note that in the current version of the \fermi\ Science Tools, an unbinned likelihood analysis of a faint source in the Galactic Plane can lead to an overestimation of the test statistic (TS\footnote{http://fermi.gsfc.nasa.gov/ssc/data/analysis/LAT\_caveats.html}). However, this effect should have a minimal impact on our analyses since the bias increases with exposure time whereas we are concentrating on exposures between 0.1--10\,d.

\section{Results}
\label{sec_res}

\subsection{Cygnus~X-3}

Figure\,\ref{fig_lc_CygX3} presents the light curve for \cygiii\ resulting from the unbinned LAT likelihood analysis, i.e., the source photon flux integrated over the 0.1--10\,GeV energy range, its error, and a test statistic (TS $\sim -2\ln L$ where $L$ is the ratio of the likelihood of models without and with the source of interest, c.f., \citealt{mat96}). Test statistics and their corresponding significances ($\sigma \sim \sqrt{\mathrm{TS}}$) are not trial-corrected. For time bins in which TS$<$20, a 1-$\sigma$ upper limit on the flux is plotted instead.

Provided for comparison are the daily light curves in soft and hard X-ray bands from \maxi-GSC (2--4\,keV), \rxte-ASM (3--5\,keV), and \swift-BAT (15--50\,keV) \footnote{http://maxi.riken.jp \\ http://xte.mit.edu/ASM\_lc.html \\ http://heasarc.gsfc.nasa.gov/docs/swift/results/transients}. Note the significant degradation of the ASM data towards the end of the \rxte\ mission lifetime \citep[after MJD\,55200: e.g.,][]{gri13}. As can be seen in Fig.\,\ref{fig_lc_CygX3}, the gamma-ray sampling by the LAT covers a wide range of spectral states (and transitions). Table\,\ref{tab_cygx3} summarizes the daily gamma-ray detections of \cygiii\ and the corresponding X-ray count rates. References to gamma-ray detections reported within 3 days of these dates are included. 

Our analysis of \cygiii\ reproduced all previously-reported gamma-ray outbursts detected with LAT \citep{corbel09,corbel10,wil11,corbel11,corbel12} at comparable fluxes and TS values ($F_{\geq \rm{100\,MeV}} = (1-4)\times10^{-6}$\,\phcms). The source has also been detected on several occasions with \agile\ \citep{bul10b,bul11a,bul11b,bul11c,pia11,pia12}, and these are indicated in Fig.\,\ref{fig_lc_CygX3} by ticks at the top of each panel. We can confirm significant LAT detections (TS$\ge$20) around these dates for all but one of these: 2011 May 28 \citep[MJD\,55709:][]{pia11} which happens to be the last reported gamma-ray outburst from this source. 

There are 5 new days not previously reported with either \fermi\ or with \agile\ during which the flux from \cygiii\ is detected at a significant level by the LAT:  these are MJD\,54708, 54911, 55092, 55197, and 55890 (Table\,\ref{tab_cygx3}). The light curve binned at 10 d also shows significant gamma-ray emission (TS$\geq$20) for bins that overlap these new daily outbursts. 

\begin{figure}[!t] \centering
\includegraphics[width=0.5\textwidth,angle=0]{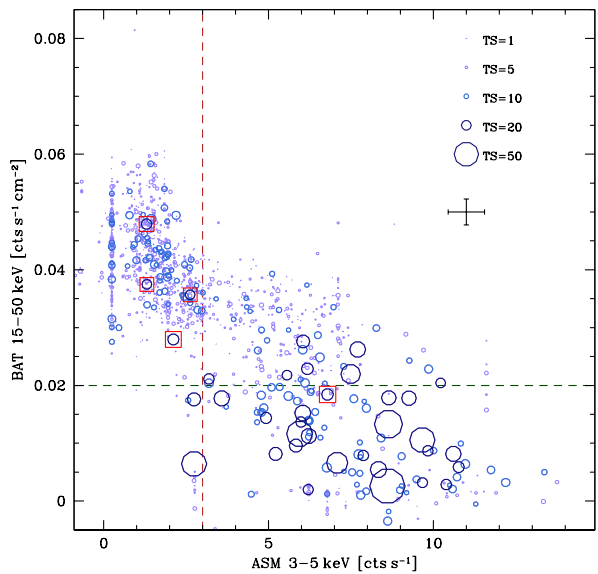}
\caption{Distribution of daily TS values (0.1--10\,GeV) for Cyg~X-3 in the ASM vs. BAT count-rate plane. The radius of each circle is proportional to the TS measured on that day, and it is centered on the reported ASM (3--5\,keV) and BAT (15--50\,keV) count rates for that day when they are available (or on the nearest date). The average 1$\sigma$ uncertainty in both directions is shown as a cross. Days with TS$\geq$20 are plotted with dark symbols, while the 5 new daily detections are boxed. The vertical dashed line indicates a count rate of 3\,\cps\ in the ASM band while the horizontal dashed line denotes a count rate of 0.02\,\cps\ in the BAT band. }
\label{fig_lc_CygX3_lat_asm_bat}
\end{figure}

Another way to visualize the X-ray and gamma-ray data is shown in Fig.\,\ref{fig_lc_CygX3_lat_asm_bat} where the distribution of daily TS values obtained for \cygiii\ is plotted in the ASM vs. BAT count-rate plane. Each circle in this plot is proportional in size to the daily TS value obtained for \cygiii\ and is centered at the reported ASM and BAT count rates on that day. In cases where either ASM or BAT did not provide a daily count rate, we used the values from the nearest date so as to not discard information. Occasionally, this leads to vertical streaks in the distribution. For example, days from the last $\sim$6 months of LAT data used in this analysis were gathered when \rxte\ was no longer operating so they share the last reported ASM count rate of 0.245\,\cps. The average error on the count rate in both X-ray bands is denoted by a cross.

Visible is the expected hard (BAT) to soft (ASM) X-ray anti-correlation despite the large scatter in the data. Days for which \cygiii\ presented a high TS value (large circles) tend to be located in the lower right portion of the plot; in the parameter space defined by \citet{corbel12} that appear necessary for gamma-ray emission from \cygiii. This is when \cygiii\ emits over 3\,\cps\ in the ASM band (3--5\,keV), and crosses the critical value of 0.02\,\cps\ in the BAT band (15--50\,keV). A multiplicative factor of $\sim$0.05 is required to convert count rates from ASM to \maxi\ (2--4 keV).

These X-ray conditions from \citet{corbel12} were met or close to being satisfied on almost all days for which TS$\geq$20, possibly meeting them on timescales less than the 1-d binning used in the light curve. On the other hand, the X-ray conditions are often satisfied without a corresponding daily LAT detection; this is evident in the large number of low-TS value circles located within this region.

\begin{figure}[!t] \centering
\includegraphics[width=0.5\textwidth,angle=0]{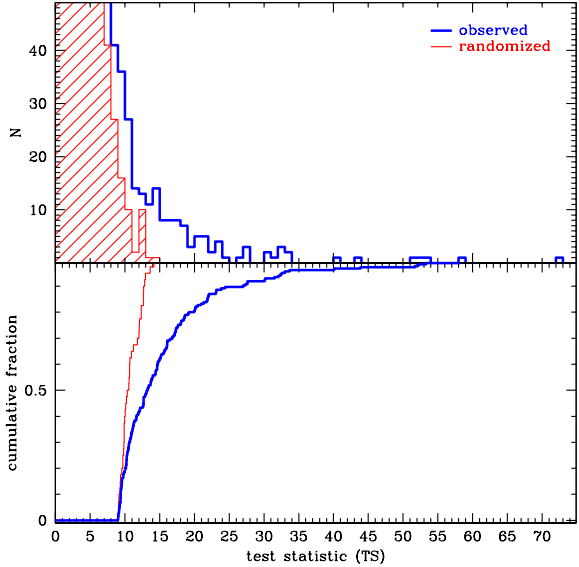}
\caption{Distributions of the test statistic (TS) from an unbinned likelihood analysis of daily LAT data of Cyg~X-3 (observed), compared with those from a spurious unidentified gamma-ray source that follows a randomized $\chi_{3}^{2}$ distribution (randomized). Restricting the cumulative distributions to TS$\ge$9 and applying a KS-test yields a low probability ($<10^{-10}$) of statistical compatibility.}
\label{fig_kstest_CygX3}
\end{figure}

In Fig.\,\ref{fig_kstest_CygX3}, we present the distribution of daily TS values from \cygiii\ as compared with a distribution of TS values that follows a randomized $\chi_{3}^{2}$ distribution \citep[i.e., the expected TS distribution for a spurious unidentified gamma-ray source:][]{mat96}. We are mainly interested in TS values of 9 or above, and the distributions diverge below this value, so we drew the cumulative distributions of the samples for TS$\ge$9 and applied the Kolmogorov-Smirnov (KS) test. We obtain a very low KS-test probability (less than $10^{-10}$) of statistical compatibility between the observed and randomly-generated distributions.

\begin{figure}[!t] \centering
\includegraphics[width=0.5\textwidth,angle=0]{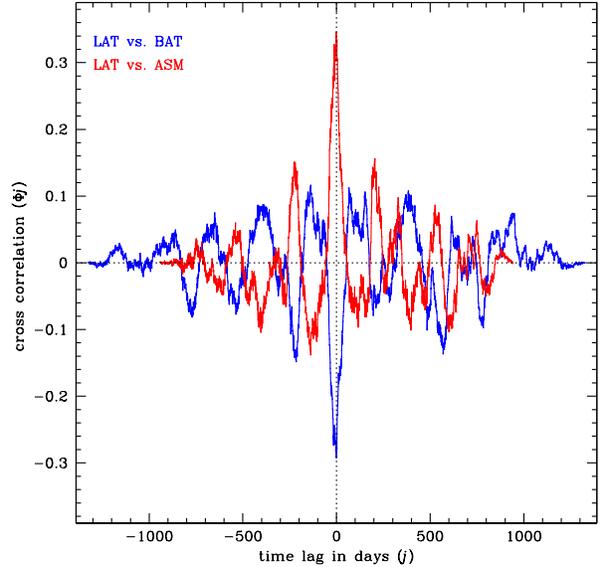}
\caption{Discrete cross-correlation functions for Cyg~X-3 comparing daily LAT (0.1--10\,GeV) light curve data with those from ASM (3--5\,keV: solid curve) and BAT (15--50\,keV: dashed curve). }
\label{fig_lc_cc_cygx3}
\end{figure}

Figure\,\ref{fig_lc_cc_cygx3} shows the discrete cross-correlation function of \cygiii\ comparing the light curve in gamma-rays with those from soft (ASM) and hard (BAT) X-rays. Deviations from cross-correlation $\phi_{j} = 0$ are $\lesssim$35\% and are likely the result of instrumental effects such as the 57-d precession period of the \fermi\ space telescope (and other known \fermi\ periods such as at 96\,min, 3.2\,hr, 1\,d and 91\,d: R.H.D. Corbet, private communication). 

It has been proposed by \citet{corbel09} that gamma-ray emission from \cygiii\ is maximized at superior conjunction, i.e., when the compact object and its jet are irradiated by the WR star with respect to our line of sight. We checked whether the 5 new daily detections of \cygiii\ occurred at specific binary orbital phases by using the updated ephemerides of \citet{zdz12} to assign the orbital phase to the events file. Merging the events files from the 5 new daily detections into a single file enables us to visually confirm the 96\,min and 3.2\,hr periodicities which are due to the orbital period of the \fermi\ spacecraft. Because these periodicities are strong compared to the emission from \cygiii, we are unable to measure a significant increase in events coincident with certain phases of the expected 4.8-hr binary orbital period. This is true even when we focus on the day during which the TS is highest (MJD\,54812). The binary period is also absent from the 0.1-d light curve generated from a likelihood analysis of the full data set which yields a significant detection only at the 57-d \fermi\ precession period. On the other hand, orbital modulations in the gamma-rays have only been detected during gamma-ray flaring epochs \citep{corbel12}, and the likelihood analysis is not recommended for exposures shorter than $\sim$1 day, so other methods such as aperture-restricted event-weighting and epoch-specific analyses \citep[e.g., as in][]{corbet11,corbel12} are more appropriate. 

Figure\,\ref{fig_lc_CygX3} showed good overall consistency between the 1-d and 10-d LAT light curves. In addition, the 10-d binned light curve reveals numerous significant detections (TS$\ge$20) of \cygiii\ both in and out of the expected gamma-ray active states. This could be the signature of faint, persistent gamma-ray emission from \cygiii, but another possibility is that these are contaminating photons originating from \psr\ which is located around 30$^{\prime}$ away. Unlike in previous studies \citep{corbel09,wil11,corbel12}, in our likelihood analysis of \cygiii, events corresponding to the on-pulse phases of this nearby pulsar were not removed. 

To test the degree to which \psr\ contributes to the gamma-ray flux at the location of \cygiii, we restricted the event times to an interval in which gamma-ray photons from \cygiii\ have been confirmed by the LAT and by \agile: between MJD\,54990 and MJD\,55045. Phase information from the latest gamma-ray ephemeris of \psr\ from \citet{ray11} was added to the events file of this time interval, and a new events file was generated in which the on-pulse phases were excluded. These are phases $0.0 \leq \Phi \leq 0.15$, $0.5\leq \Phi \leq 0.6$, and $0.95\leq \Phi \leq 1.0$ (c.f., Fig.\,52 of \citet{ray11}). This removed 30\% of the live-time and so the exposure time of the resulting events file was adjusted accordingly. We ran the likelihood analysis considering the full 55-d interval as a single observation. In this case, the spectral parameters of sources located within 3$^{\circ}$ of \cygiii\ (including \psr\ and the diffuse emission components) were allowed to vary, while normalizations were variable for sources out to a larger radius (9$^{\circ}$) around \cygiii. 

Discarding the on-pulse times removed nearly all photons from \psr, and it leads to a 30\% reduction in the photon counts from \cygiii\ and a similar reduction in the sum total of the photon counts from all sources in the field. This is consistent with the number of photons that would have been counted in the amount of exposure time that was removed. In conclusion, \psr\ is not contributing significant numbers of unaccounted-for photons at the position of \cygiii\ which makes it unlikely to be the source of persistent emission.

We can not completely rule out contaminating emission from other sources in the field, including the diffuse galactic and extragalactic background, as the source of the faint, persistent emission detected at the location of \cygiii. However, the bulk of the emission from other field sources is accounted for in the spectral models provided by the LAT team, and especially in the refined spectral models that we generated from the binned analysis of the long-term dataset. As we show in the next section, this underlying emission is not present in the 10-d binned light curve of \cygi\ which is also located in this field. This suggests that the 10-d detections in and out of previously-reported gamma-active epochs are likely to be from \cygiii\ which means that persistent gamma-ray emission is being detected from a $\mu$QSO for the first time.

\begin{figure*}[!t] \centering
\includegraphics[width=16cm,angle=0]{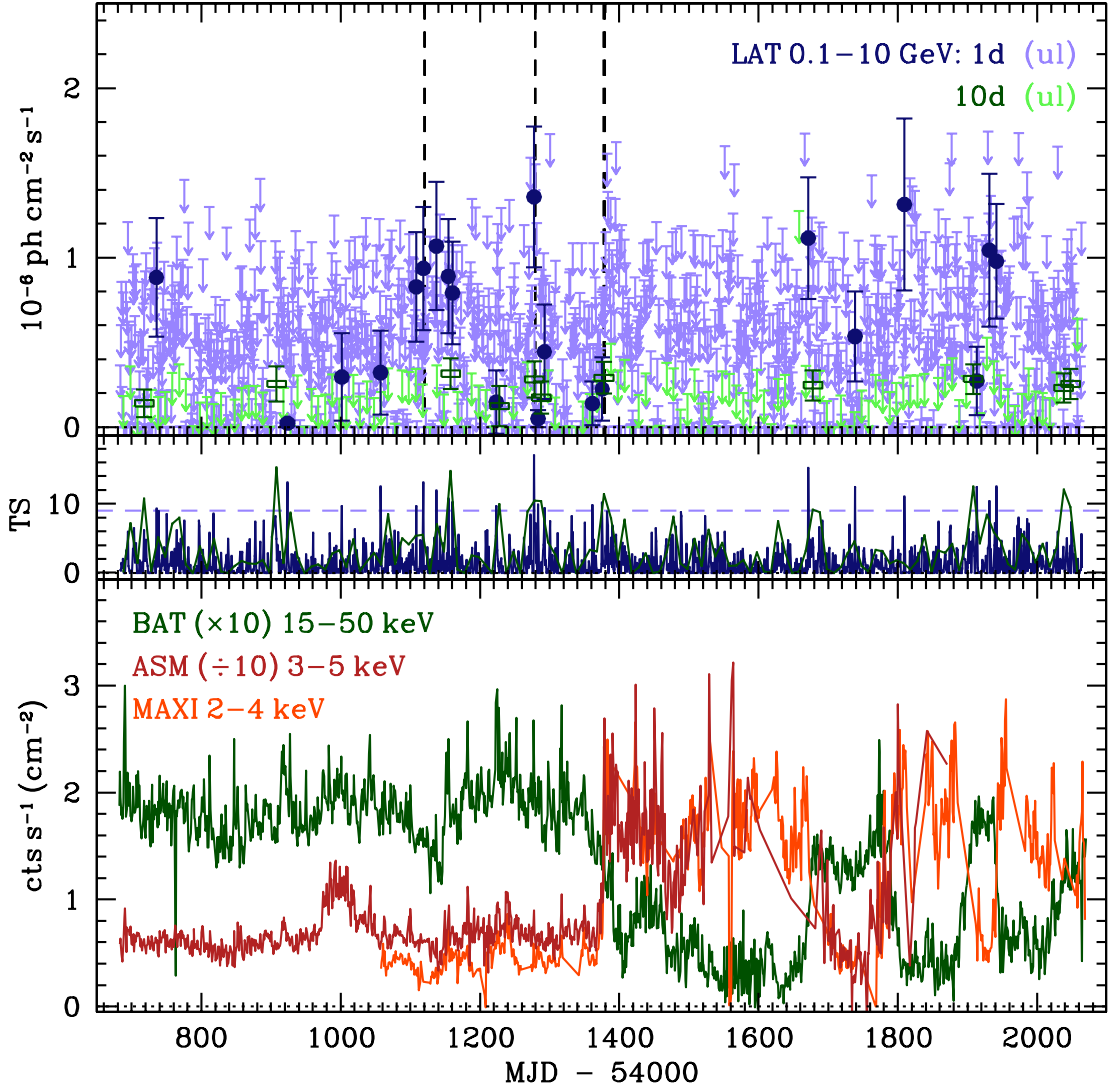}
\caption{Same as Fig.\,\ref{fig_lc_CygX3} for Cyg~X-1 with a TS threshold of 9 (horizontal dashed line in the middle panel). }
\label{fig_lc_CygX1}
\end{figure*}

\subsection{Cygnus~X-1}

In the 1369 days of gamma-ray monitoring by the LAT (Fig.\,\ref{fig_lc_CygX1}), the TS value reached a level of 9 ($\sim3\sigma$) or above on the 21 days listed in Table\,\ref{tab_cygx1}. Three of these candidate gamma-ray detections of \cygi\ by the LAT are contemporaneous with possible gamma-ray detections reported in the literature by \agile. In all three cases, the possible 1-d detections by LAT precede the detections reported with \agile\ by 1--2 days.

On 2009 October 14 (MJD\,55118), we obtained a candidate detection of \cygi\ with the LAT at a TS value of 13 ($\sim3.6\sigma$). This was only 1--2 days before \citet{sab10a} reported a short (1-d long) outburst at the location of \cygi\ with \agile\ \citep[see also][]{sab13}. With the LAT, we measured a source flux of $(1.1\pm0.4)\times10^{-6}$\,\phcms\ in 0.1--10\,GeV, i.e., slightly lower than, but statistically compatible with, the flux reported with \agile.

\begin{figure}[!t] \centering
\includegraphics[width=0.45\textwidth,angle=0]{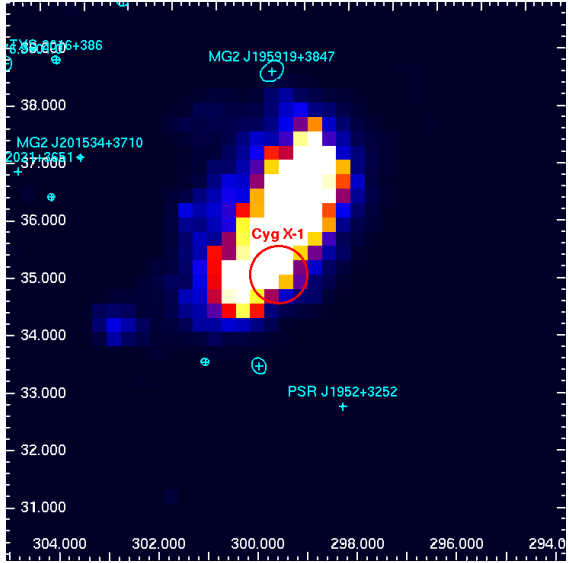}
\caption{Differential test statistic (TS) map of a $10^{\circ}\times10^{\circ}$ field centered on Cyg~X-1 from LAT data (0.1--10\,GeV) during MJD\,55277. Each pixel is 0.25$^{\circ}$ wide, and in the lightest regions, the TS values range from 15 to 17 ($\sim4\sigma$).}
\label{fig_CygX1_tsmap}
\end{figure}

The maximum value of TS$=$17 ($\sim4\sigma$) was reached on 2010 March 22 (MJD\,55277), i.e., 1--2 days before a gamma-ray flare was detected by \agile\ \citep{bul10a}. On this date, LAT recorded a source flux of $(1.4\pm0.4)\times10^{-6}$\,\phcms\ (0.1--10\,GeV), i.e., consistent with the flux reported by \agile\ in a similar band: $(2.0\pm0.9)\times10^{-6}$\,\phcms. Figure\,\ref{fig_CygX1_tsmap} presents a spatial map of the TS value on MJD\,55277 generated by modeling all sources except for \cygi. A bright cluster of pixels can be seen consistent with the location of \cygi. The large size of this pixel cluster and its offset with respect to \cygi, both of which are also seen in \agile\ TS maps of this source \citep{sab13}, can be explained by the $\sim$2$^{\circ}$-diameter of the PSF at these energies and the low photon counts. 

Around the time of the third \agile\ detection \citep[beginning on 2010 June 30 or MJD\,55377:][]{sab10b,sab13}, we obtained with LAT on 2010 June 28 (MJD\,55375) a TS value of 10 ($\sim3\sigma$) at a flux of $2\times10^{-7}$\,\phcms, i.e., less than the flux of $10^{-6}$\,\phcms measured with \agile. Figure\,\ref{fig_lc_CygX1} shows that this date corresponds to an epoch in which \cygi\ was undergoing a hard-to-soft state transition \citep{neg10,rus10,wil10}. There are no reports of contemporaneous radio flaring which is sometimes seen during such transitions \citep[e.g.][]{cad06,wil07}. Monitoring around this time shows that the radio emission was briefly quenched \citep[MJD\,55386:][]{rus12} and then showed the possible spectral signatures of discrete ejection events \citep[MJD\,55400:][]{tud10}.

\begin{figure}[!t] \centering
\includegraphics[width=0.5\textwidth,angle=0]{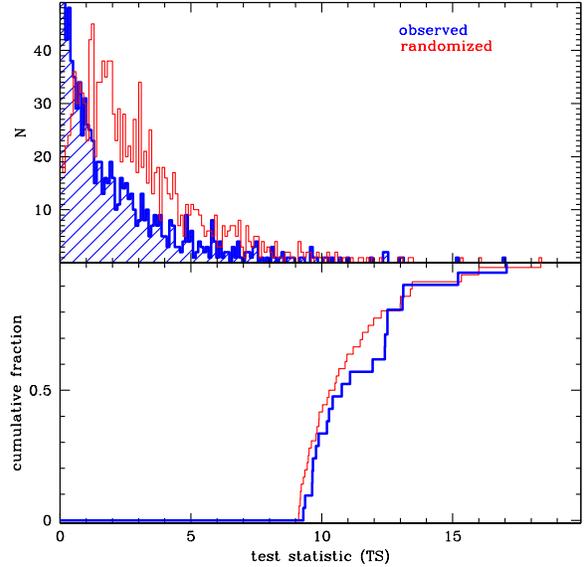}
\caption{Same as Fig.\,\ref{fig_kstest_CygX3} for Cyg~X-1. For TS$\ge$9, the probability of statistical compatibility from a KS-test is 55\%.}
\label{fig_kstest_CygX1}
\end{figure}

Given that there are contemporaneous (but not coincident) detections with \agile\ for MJD\,55118, MJD\,55277, and MJD\,55375, it is not necessary to apply a trial correction to the detection significances that we obtained on these dates. For the other days listed in Table\,\ref{tab_cygx1}, the requirement to correct for the number of blind-search trials can be relaxed if it can be shown that these dates correspond to likely gamma-ray emission states from \cygi, e.g., states in which the jet is present. The intermediate state of \cygi\ is thought to feature the formation and destruction of relativistic jets, while compact jets (lower bulk velocity, and hence, less energetic than the relativistic jets) are believed to always be present during the hard state \citep{sti01,fen09}.

We identified the most likely state of \cygi\ by employing the prescription of \citet{gri13} based on \rxte\ and \emph{MAXI} count rates. There are 5 days listed in Table\,\ref{tab_cygx1} in which \cygi\ was in the transitional/intermediate state (MJD\,55001, 55160, 55671, 55913, and 55941), and 1 day (MJD\,55809) where the source was in the soft state, which suggests that it passed through the intermediate state on timescales shorter than the binning that was used. The other dates in Table\,\ref{tab_cygx1} had \cygi\ in the hard state.

As was done for \cygiii, we compared the observed TS distribution from \cygi\ with that of a randomly-generated light curve of a spurious unidentified EGRET source (Fig.\,\ref{fig_kstest_CygX1}). For TS$\ge$9, the KS-test probability of statistical compatibility is 55\%. This suggests that low-significance LAT detections that are not contemporaneous with those of another gamma-ray instrument (e.g., \agile), or for which there is no other compelling evidence that gamma-rays should be present, are likely spurious. Signatures of the 5.6-d binary orbital period of \cygi\ \citep[e.g.,][]{wen06} could not be found in the 0.1-d light curve.

\begin{figure*}[!t] \centering
\includegraphics[width=16cm,angle=0]{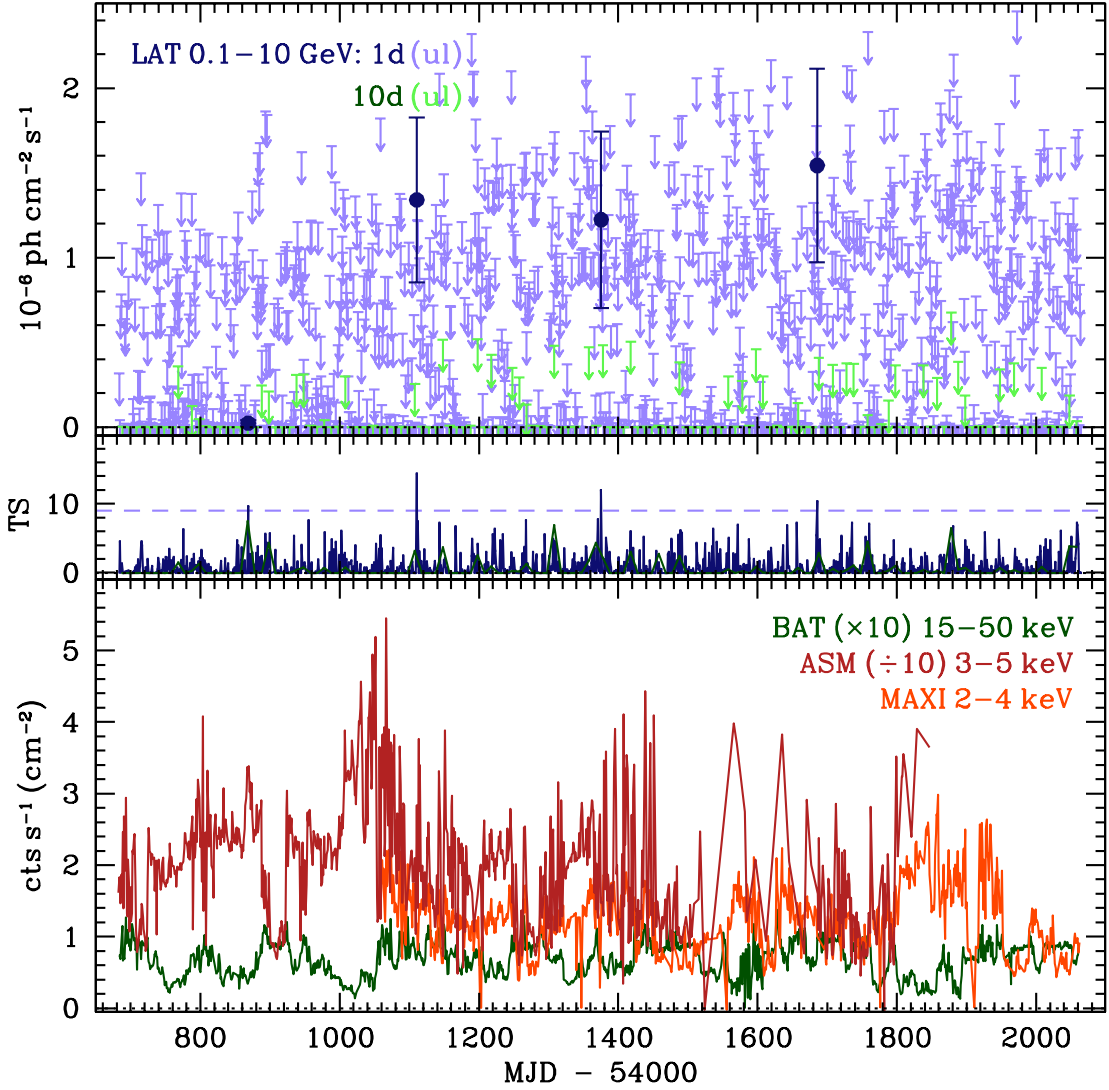}
\caption{Same as Fig.\,\ref{fig_lc_CygX1} for GRS~1915+105.}
\label{fig_lc_grs1915}
\end{figure*}

\subsection{GRS~1915$+$105}

\begin{figure}[!t] \centering
\includegraphics[width=0.5\textwidth,angle=0]{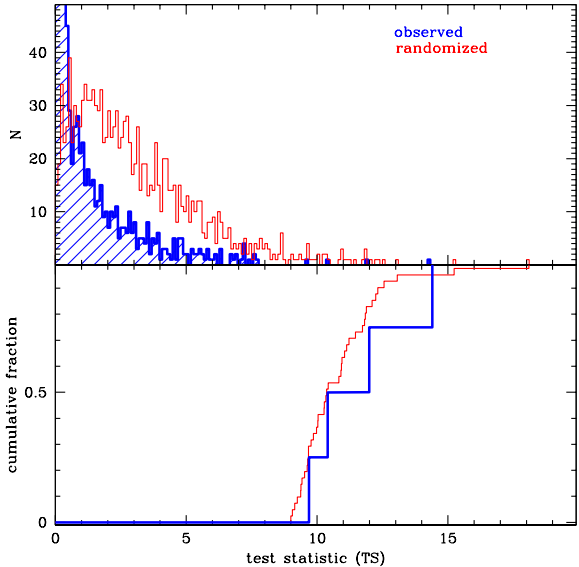}
\caption{Same as Fig.\,\ref{fig_kstest_CygX3} for GRS~1915+105. The KS-test probability is 83\%.}
\label{fig_kstest_grs1915}
\end{figure}

The light curve of \grs\ from the likelihood analysis of 1-d and 10-d bins contains four days (out of 1371) on which the TS was above 9 (a maximum of 14.4 was reached on MJD\,55110) with no 10-d bins having TS$\ge$9 (Fig.\,\ref{fig_lc_grs1915} and Table\,\ref{tab_grs1915}). Comparing the observed daily TS distribution with that of a randomized TS distribution yields a KS-test probability of 83\% of statistical compatibility between the distributions for TS$\ge$9 (Fig.\,\ref{fig_kstest_grs1915}). This indicates that the low-significance detections are likely spurious. It is also clear that the source is not detectable on longer timescales of 1--2 years based on its absence from the 1FGL and 2FGL catalogs. We obtain a 3$\sigma$ upper limit of $2.3\times10^{-8}$\,\phcms\ on the persistent flux from \grs\ in the 0.1--10\,GeV range. This value was determined by considering the full 4-year data set as a single observation, and running the likelihood analysis where the spectral parameters of sources within 3$^{\circ}$ of \grs\ (including those of \grs\ and of the diffuse galactic and extragalactic components) were allowed to vary, while sources within 9$^{\circ}$ of \grs\ had free normalizations.

\subsection{GX~339$-$4}

\begin{figure*}[!t] \centering
\includegraphics[width=\textwidth,angle=0]{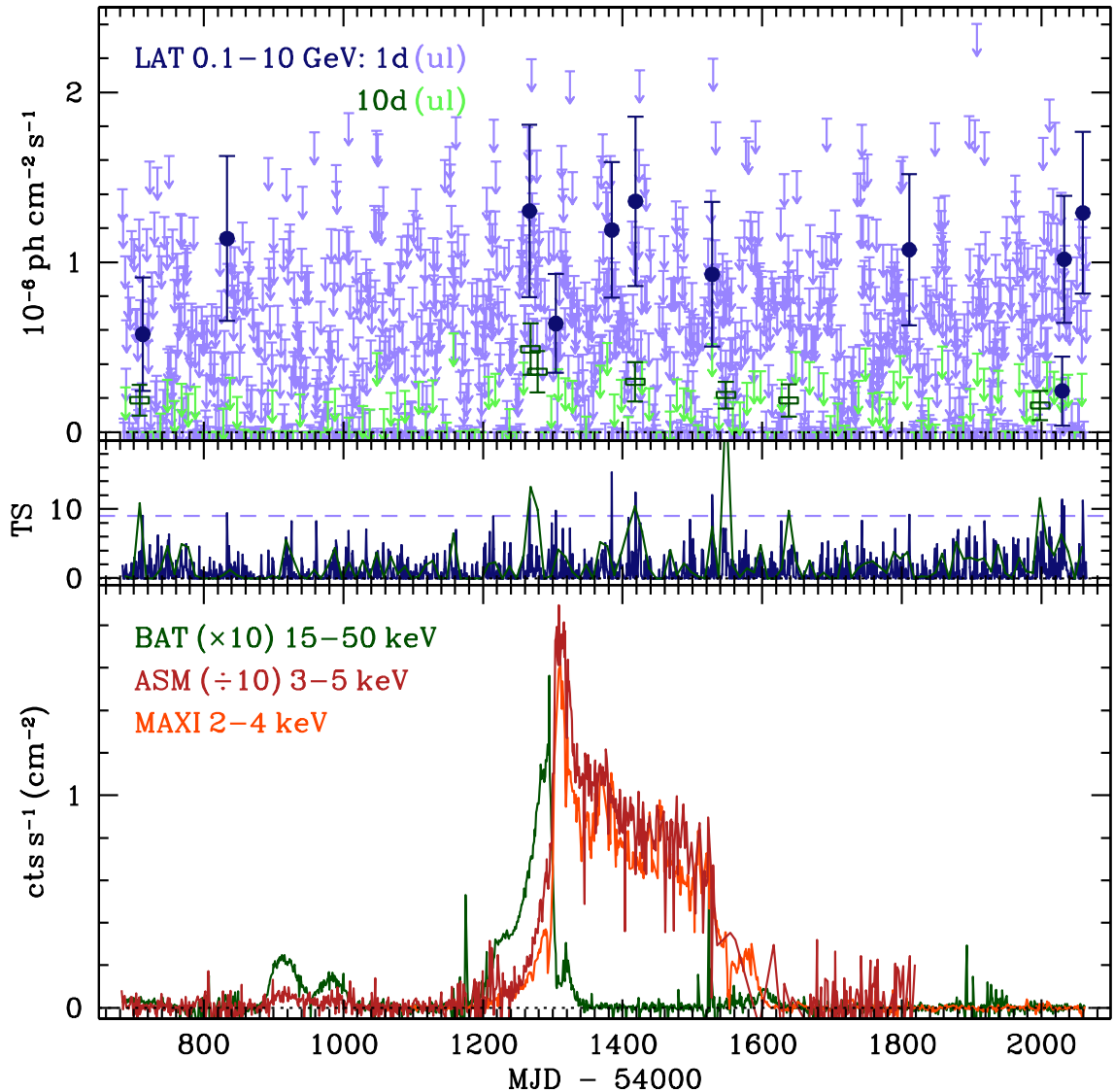}
\caption{Same as Fig.\,\ref{fig_lc_CygX1} for GX~339$-$4.}
\label{fig_lc_gx339}
\end{figure*}

\begin{figure}[!t] \centering
\includegraphics[width=0.5\textwidth,angle=0]{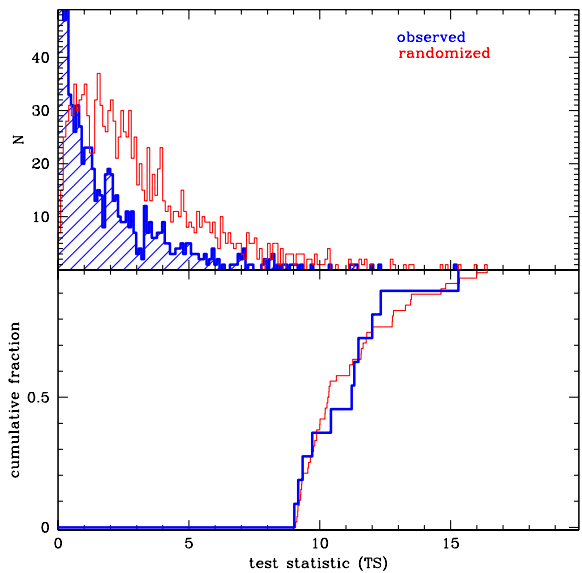}
\caption{Same as Fig.\,\ref{fig_kstest_CygX3} for GX~339$-$4. The KS-test probability is 73\%.}
\label{fig_kstest_gx339}
\end{figure}

There are hints that the large X-ray outburst of 2010 was accompanied by TS$\ge$9 detections on 1-d and 10-d timescales in the gamma-rays (Fig.\,\ref{fig_lc_gx339}). This can be seen in the apparent clustering of TS$\ge9$ bins within the $\sim$400 day duration of the X-ray flare. However, the significances in these bins are low ($\lesssim$4.5$\sigma$) and so the data do not provide conclusive evidence of gamma-ray activity by \gx\ on 1-d or 10-d timescales in the nearly 4 years since the launch of \fermi. The known orbital period of 1.8\,d \citep{jon04} is not detected in the 0.1-d light curve. We can also rule out a detection on timescales of years given its non-inclusion in the 1FGL and 2FGL catalogs. This is despite the fact that the source has shown multiple state transitions during this time, including the brightest radio emission ever detected from this source during the hard state \citep{corbel10b}, and the ejection of material along a relativistic jet colliding with the interstellar medium \citep{corbel10c}. On 11 days (out of 1369) the TS was at the level of 9 or above (Table\,\ref{tab_gx339}). The KS-test probability of 73\% suggests that the low-significance detections are likely spurious (Fig.\,\ref{fig_kstest_gx339}). Based on a likelihood analysis of the 4-year data set in which the spectral parameters of \gx\ and nearby sources (including those of the galactic and extragalactic diffuse background emission) were allowed to vary, we obtain a 3$\sigma$ upper limit of 1.6$\times10^{-8}$\,\phcms\ on the persistent 0.1--10-GeV flux from \gx.

\section{Discussion}
\label{sec_disc}

\subsection{Gamma-ray emission from Cygnus~X-3}

Our analysis of \cygiii\ produced results that are consistent with previous reports and provided additional insight into this $\mu$QSO. We confirm all but one of the previously-reported gamma-ray detections at similar flux levels and significances. The lone exception is the last reported gamma-ray detection by \agile\ \citep{pia11}. 

Five new transient emission events (lasting $\sim$1 day) were detected outside of any previously-known gamma-ray active epochs. Two of the five events satisfy the X-ray conditions necessary for triggering gamma-ray emission as defined by \citet{corbel12}. A third criterion relating to radio emission also exists: in past gamma-ray detections, the radio emission was either quenched just prior to the appearance of gamma-rays, or it showed minor flaring with flux densities of $\sim$0.1--0.6\,Jy at 15\,GHz. We checked the radio emission at 15\,GHz on these dates with the Large Array of the Arcminute Microkelvin Imager \citep{zwa08}, and found that the radio emission was nearly quenched (flux density $\sim$10\,mJy) on MJD\,55092. For the other 4 candidate gamma-ray detections, the radio flux density varied between 0.05\,Jy and 0.15\,Jy, i.e., at the low end of the level of minor flaring that has been seen in previous gamma-ray outbursts.

One observable difference between leptonic and hadronic processes is the cutoff energy in the gamma-ray range of the spectral energy distribution (SED). In IC processes, the electrons can lose energy very rapidly by scattering and so they are unable to reach energies higher than a few TeV. Protons are not able to cool as efficiently and so the cutoff of the SED in the hadronic case extends to higher energies than in the leptonic case. The source flux and upper limits are not constraining enough to enable us to conclusively confirm or reject one of these scenarios.

Another observable difference between leptonic and hadronic processes is that neutrinos are the expected byproduct (besides gamma-rays) of the decay of neutral pions. Detecting neutrinos from \cygiii\ would provide conclusive evidence that hadronic processes dominate the high-energy range of the SED \citep{chr06,bae12}. Observations gathered with IceCube \citep{ahr03} have not yielded a signature of a point-like neutrino source at the position of \cygiii, however, the expected neutrino flux is well below the current sensitivity limit of the instrument, and so more data are needed \citep{abb11}.

We detected significant ($\gtrsim5\sigma$ pre-trial correction) gamma-ray emission from \cygiii\ on timescales of 10 days or more. Many of these 10-day-long bins overlap days on which transient outbursts were reported above $\sim$100\,MeV. These outbursts occasionally last a few days or weeks, so detecting emission in 10-d bins that are coincident with these epochs is not surprising. However, we did not anticipate finding a significant number of 10-d bins with high TS values outside of these gamma-active epochs, i.e., no more than would be expected from a likelihood analysis of a spurious unidentified gamma-ray source. Instead, we found that around half of the bins on the 10-d timescale had TS$\ge$20 ($\gtrsim4.5\sigma$), and many of these were outside any known gamma-ray active epoch. 

Contaminating emission from point-like or diffuse sources near \cygiii, which happens to reside in one of the most challenging regions of the sky for the LAT at these energies, could not be completely excluded. \citet{ner12} found that the pulsar population contributes significant variability (on timescales of months) to the gamma-ray emission along the Galactic Plane. There is still a great deal of uncertainty in the spectral and spatial models of the sources in the field, and this is especially true for the diffuse components. Nevertheless, these models represent the best fit available over the full data set. If we assume that the underlying emission is not from \cygiii\ but due to the summed contribution of pulsars and other sources, this indicates that the normalizations of sources in the Cygnus region should be slightly higher (within the uncertainties) than they are currently set in the 2FGL catalog. This hypothesis can be rejected based on the fact that this emission is not present in the 10-d light curve of another object in this field with similar source spectral models: \cygi. 

If this is really coming from \cygiii, then it appears to be persistent gamma-ray emission. It might be that this system is unique given its extremely tight orbit and very massive donor star. The dense wind of the WR star being in such close proximity to an accreting CO can lead to shocks and particle acceleration inside turbulent accretion zones \citep{bed09}. \citet{fen09} propose that steady, compact jets are expected from accreting black holes in the hard state. While \cygiii\ was often observed to be in the hard state during the past 4 years, it is important to note that interactions between the accretion disk and the wind from the WR donor star lead to significant differences in the hard state spectrum of this source compared with the typical hard state of BH X-ray binaries \citep{szo08a,szo08b}. Another possibility is that the jet in \cygiii\ never turns off, and that these jet electrons IC-scatter photons from the dense radiation field \citep{der06}. If the plasma around the black hole is pair-dominated or mildly relativistic, then it can generate a gamma-ray ``bump'' at MeV energies, although not persistently \citep{lia88,der96,li96,mcc00}.

\subsection{Possible confirmation of the gamma-ray detection of Cygnus~X-1}

In the absence of any indication that gamma-rays should be present from \cygi\ at a specific epoch, we would ascribe the few dozen low-significance detection days as being spurious. However, on three occasions (MJD\,55118, MJD\,55277, and MJD\,55375), there was corroborating evidence from \agile\ that gamma-rays were contemporaneously detected from this source. Our analysis of the LAT data supports the conclusions by the \agile\ team that \cygi\ was gamma-ray active around these dates \citep{bul10a,sab10a,sab10b,sab13}. Prior analysis of the LAT data by our group, by others \citep[e.g.,][]{hil11}, and by the \agile\ team (S. Sabatini: private communication) did not result in detections on these dates, and so the new Pass 7 data used in the current analysis is the first time that such excesses have been seen with the LAT. 

In all three cases, the detection by \fermi\ preceded that of \agile\ by 1--2 days. Given the lower energy bandpass of \agile\ compared with the LAT, this delay could be an indication that different parts of the jet are producing gamma-rays at different times. For example, gamma-rays produced near the base of the jet are first seen at high energies by \fermi, and 1--2\,d later, these jet particles interact with the ISM producing lower-energy gamma-ray photons seen by \agile. Another possibility is that the delay is related to the cooling of accelerated jet particles via adiabatic expansion, or via synchrotron and brehmsstrahlung losses. 

It has been proposed (and possibly observed once with MAGIC) that the production of gamma-rays should be maximized in IC and pair-production cascades near the phase of superior conjunction, i.e., when the massive donor star is between the observer and the CO, so that relativistic particles from the jet are beamed into the soft stellar radiation field \citep[e.g.][]{bed97,bed07,rom10, dub10}. In \cygi, the time of superior conjunction in MJD is $T_{0} = 50077.995 + nP$ where $n$ is the number of orbital cycles and $P$ is the orbital period of 5.599829(16)\,d \citep{bro99}. 

The first LAT detection of \cygi\ contemporaneous with an \agile\ detection was on MJD\,55118. This integration window begins only 0.03 in phase after the nearest superior conjunction at MJD\,55117.84(1) ($n=900$). The second LAT detection contemporaneous with an \agile\ detection (MJD\,55277) is in the phase range 0.40--0.58 which overlaps inferior conjunction, while the \agile\ integration window begins within 0.03 in phase of superior conjunction. The third LAT detection contemporaneous with an \agile\ detection (MJD\,55375) completely covers superior conjunction at MJD\,55375.43(1) ($n=946$). Keep in mind, however, that the short orbital period implies a $\sim$20\% probability that superior conjunction will occur on any given day. This is consistent with our observations whereby superior conjunction fell on 3 of the 21 days (15\%) for which TS$\ge$9. 

On these 21 days with TS$\ge$9, \cygi\ was around twice as likely (29\% vs. 15\%) to be in the transitional (or intermediate) state, where the conditions favor the formation and destruction of the relativistic jet. Both of the contemporaneous candidate gamma-ray detections by \agile\ and \fermi-LAT were captured while the source was in the hard state, where a steady, low-velocity, compact jet is expected \citep{sti01}. During the review of this article, \citet{mal13} reported a detection of \cygi\ with the LAT (TS $=$ 15.6, $\sigma \sim 4$) by summing all available gamma-ray data during the hard state. The radio data from AMI shows \cygi\ at a flux density of 5--20\,mJy at 15\,GHz on these 2 days (and similar flux densities around the other 19 days) with no correlation between the level of radio emission and the significance of the candidate gamma-ray detections. 

The data do not allow us to paint a complete gamma-ray picture of \cygi, but it is clearly a candidate gamma-ray binary that will continue to merit further observations and study in the high-energy domain.

\subsection{Ingredients necessary to produce gamma-rays in $\mu$QSOs}

The recipe for producing transient gamma-rays in \cygiii\ appears to be straightforward \citep{corbel12}: variable radio emission from the jet at the level of $\sim$0.2--0.4\,Jy (15\,GHz); a soft X-ray flux in the ASM band above 3\,\cps (3--5\,keV); and a hard X-ray flux in the BAT that transitions through 0.02\,\cps\,\cmsq\ (15--50\,keV). These LAT observations provide us with a more general list of ingredients necessary for gamma-ray production in $\mu$QSOs.

According to leptonic and hadronic emission models, the jet is the primary ingredient necessary for producing gamma-rays in $\mu$QSOs. These jets appear during specific X-ray emission states and so we have a direct relation between the soft vs. hard accretion-powered X-ray luminosity, accelerated particles and synchrotron emission from the jet, and the production of gamma-rays. Searches for gamma-rays from these systems should focus on epochs with intense radio flaring. High-energy photons originating within the $\mu$QSO can be photo-absorbed by the various radiation fields that are present \citep{aha04,bos08,dub10}. This absorption tends to dominate at energies $\gtrsim$10\,GeV, i.e., above the energy range considered in our analysis.

Only one $\mu$QSO has been positively identified in the gamma-rays: \cygiii. There are possible (low-significance) detections of \cygi\ in the gamma-rays with MAGIC \citep{alb07}, with \agile\ \citep{bul10a,sab10a,sab10b,sab13}, and now with \fermi. Thus, the two $\mu$QSOs with confirmed or possible gamma-ray detections are both HMXBs. In fact, all gamma-ray emitting binaries that have been detected thus far have a high-mass companion: this includes \lsi, \lsii, \object{1FGL\,J1018.6$-$5856} \citep{ack12}, {PSR~B1259$-$63} \citep[e.g.,][]{tam11}, and possibly \object{HESS~J0632+057} \citep{hin09}. There are no known gamma-ray emitting LMXBs including, as was shown here, \grs\ and \gx. 

This suggests that in addition to the radio jets, the nature of the donor star is an important ingredient in producing gamma-rays. High-mass stars feature a hot radiation field and dense winds which can serve: as soft seed photons necessary for pair creation and IC processes; and as a medium containing photons and protons with which the jet particles can interact. This ingredient is lacking in \grs\ and \gx, which could explain why they were not detected in the gamma-rays despite the fact that both sources underwent major state transitions, including a huge radio outburst and relativistic ejection event from \gx, during the 4-year monitoring period by the LAT. Thus, there are three possibilities for LMXB microquasars: 1) they do not emit any gamma-rays at all; 2) they emit some gamma-rays, but the flux is below the sensitivity limit of current telescopes; and 3) they emit gamma-rays, and the flux should be detectable. Scenarios 1 and 2 remain valid, but scenario 3 appears to be unlikely given our results (and those of other groups).

Leptonic and hadronic models predict orbital variations of the gamma-rays since the power of the collimated jet and the photon density of the soft radiation field will change along the orbit. As a result, the binary's orbital configuration with respect to our line of sight strongly influences the observed gamma-ray spectrum from photon absorption or IC processes \citep[e.g.][]{bed97,bed07,rom10}. For \cygiii, \citet{corbel09} determined from the orbital modulations that its gamma-rays peak whenever the base of the jet is illuminated by the stellar UV photons with respect to our line of sight, i.e., near superior conjunction \citep{corbel09}. For \cygi, the possible MAGIC detection of TeV emission occurred near superior conjunction (i.e., with the CO behind the massive donor) when gamma-ray opacity is highest while the supergiant star provides a large target of UV seed photons \citep{alb07,rom10}. Superior conjunction fell within (or nearly within) the 1-d \fermi\ integration window on two of three occasions for which there were contemporaneous detections of \cygi\ by \fermi-LAT and \agile. 

High magnetic fields can suppress pair-creation cascades and lead to modulation of gamma-rays from IC scattering \citep[e.g.,][]{par08}. Jet properties are not fully known for these systems, but their inclination with respect to the accretion disk, the observer, or the material at the termination shocks should influence the gamma-ray emission. For example, \cygi\ has a super-orbital period of $\sim$300\,d \citep[e.g.,][]{zdz11} due to the precession of the accretion disk, although this variability timescale could not be detected in our gamma-ray data.

In summary, the ingredients that seem necessary for producing gamma-rays in $\mu$QSOs are: 1) a radio jet; 2) dense winds and strong radiation fields typical of a high-mass donor star; and 3) an orbital configuration in which the CO is near superior conjunction. The magnetic field within the system and the jet inclination angle are also expected to play a role. The jet should be considered the primary ingredient in gamma-ray production, but our findings indicate, as others have suggested, that the nature of the donor star is also important. If the persistent emission that we detected from \cygiii\ is confirmed, then this system, with its unique blend of extreme wind and orbital characteristics, may not even require a relativistic jet component to produce gamma-rays, although a compact jet could still be present.

\section{Summary \& Conclusions}
\label{sec_conc}

In this work, we presented the results from a systematic search for gamma-ray emission from four $\mu$QSOs using $\sim$4 years of \fermi-LAT observations in the 0.1--10\,GeV range. All four targets were very active in the X-rays during this time so our gamma-ray monitoring sampled a variety of states and transitions.

Gamma-ray flares from \cygiii\ were previously seen by \fermi\ and by \agile, and our analysis reproduced all but one of these reported outbursts at similar fluxes and detection significances. \cygiii\ was also detected at (pre-trial-corrected) significances of $\sim$5$\sigma$ on five additional days not previously reported in the literature. The known outbursts, and the five new ones reported here, all coincided with epochs in which the soft and hard X-ray fluxes satisfied (or were close to satisfying) the criteria defined by \citet{corbel12}, possibly satisfying the criteria on timescales less than the binning used here. Significant detections were seen for 10-d bins that overlap the known and new daily outbursts. In addition, there are numerous significant detections on 10-d timescales outside known gamma-ray flaring epochs. Its origin would therefore be unrelated to the relativistic jet, although a steady compact jet scenario is still a possibility, and would suggest that persistent gamma-ray emission from \cygiii\ has been detected for the first time. 

For the three other $\mu$QSOs in our study, the situation is very different from that of \cygiii. There have been three reports of possible gamma-ray flares from \cygi: one at TeV energies by MAGIC and three others in the MeV range by \agile. Our analysis yielded 21 dozen low-significance detections on 1-d timescales from \cygi, three of which were contemporaneous (within 1--2 days) with \agile\ detections. However, we conclude that the other detections are probably spurious given the absence of compelling evidence of jet or gamma-ray activity. There are no previous reports of gamma-ray flaring by \grs\ and by \gx\ and our analysis did not uncover any significant detections of these sources on 1-d and 10-d timescales. This is despite the fact that both sources were very active in the past 4 years, with \gx\ emitting a large radio flare during our gamma-ray monitoring. 

These results enable us to refine the list of ingredients that appear necessary for the detection of gamma-rays from microquasars. These ingredients are, in order of importance: the jet; the spectral type of the donor star (i.e., the dense, ionized winds of massive stars in HMXB systems are preferred); and the orientation of the jet with respect to the observer (i.e., the preferred orbital configuration of the system is superior conjunction). Other factors such as the magnetic field and the jet inclination angle could also affect the emission of gamma-rays. 

Even with the sensitivity of current instruments, detecting gamma-rays from $\mu$QSOs is challenging due to the relative faintness of these systems and their location in the crowded regions of the Galactic Plane. Nevertheless, studying these objects in the gamma-rays, in tandem with multi-wavelength monitoring from the radio and X-rays, can provide valuable insights into the physics of jets and particle acceleration mechanisms.

\acknowledgments
The authors thank the anonymous referee whose constructive criticism led to an improved manuscript. AB thanks St\'{e}phane Corbel, Robin H.D. Corbet, Victoria Grinberg, Michael McCullough, and Andrzej Zdziarski for useful discussions. AB and JAT acknowledge partial support from NASA Fermi Guest Observer Award NNX10AP83G. This research has made use of: data obtained from the High Energy Astrophysics Science Archive Research Center (HEASARC) provided by NASA's Goddard Space Flight Center; the SIMBAD database operated at CDS, Strasbourg, France; NASA's Astrophysics Data System Bibliographic Services; the Fermi Science Support Center; the Swift/BAT transient monitor results provided by the Swift/BAT team; and MAXI data provided by RIKEN, JAXA and the MAXI team.

\bibliographystyle{apj}
\bibliography{bod.bib}
\clearpage

%
\begin{deluxetable}{ l c c c c c l }
\tabletypesize{\scriptsize}
\tablewidth{0pt}
\tablecaption{Candidate Detections of Cyg~X-3 by \fermi-LAT}
\tablehead{
\colhead{MJD} & \colhead{TS \tablenotemark{a}}  & \colhead{LAT \tablenotemark{b}}  & \colhead{ASM \tablenotemark{c}} & \colhead{BAT \tablenotemark{d}} & \colhead{criteria \tablenotemark{e}} & \colhead{reported \tablenotemark{f}} }

\startdata

54708 & 21.7 & 1.8$\pm$0.5 & 1.298$\pm$0.161 & 0.048$\pm$0.002 & N & new \\
54759 & 21.9 & 1.5$\pm$0.4 & 5.984$\pm$0.122 & 0.014$\pm$0.002 & Y$-$ & 1  \\
54768 & 27.0 & 1.9$\pm$0.5 & 6.050$\pm$0.107 & 0.028$\pm$0.003 & Y$+$ & 1, 2 \\
54804 & 21.3 & 1.7$\pm$0.4 & 6.207$\pm$0.187 & 0.002$\pm$0.001 & Y$-$ & 1 \\
54807 & 22.0 & 1.6$\pm$0.4 & 10.391$\pm$0.230 & 0.003$\pm$0.001 & Y$-$ & 1 \\
54809 & 33.7 & 1.8$\pm$0.4 & 8.334$\pm$0.258 & 0.005$\pm$0.001 & Y$-$ & 1 \\
54810 & 32.1 & 2.3$\pm$0.5 & 10.609$\pm$0.325 & 0.008$\pm$0.001 & Y$-$ & 1 \\
54811 & 21.9 & 1.6$\pm$0.5 & 9.835$\pm$0.300 & 0.009$\pm$0.002 & Y$-$ & 1 \\
54812 & 72.7 & 3.3$\pm$0.5 & 8.605$\pm$0.249 & 0.003$\pm$0.002 & Y$-$ & 1, 2 \\
54813 & 27.3 & 2.0$\pm$0.5 & 5.214$\pm$0.302 & 0.008$\pm$0.001 & Y$-$ & 1 \\
54814 & 40.1 & 2.2$\pm$0.4 & 7.491$\pm$0.249 & 0.022$\pm$0.002 & Y$+$ & 1 \\
54816 & 30.2 & 1.8$\pm$0.4 & 9.258$\pm$0.247 & 0.017$\pm$0.001 & Y$-$ & 1 \\
54911 & 23.1 & 1.6$\pm$0.4 & 2.114$\pm$0.143 & 0.028$\pm$0.003 & Y$+$ & new \\
54997 & 23.2 & 2.1$\pm$0.5 & 4.922$\pm$0.229 & 0.014$\pm$0.001 & Y$-$ & 1, 3 \\
55001 & 33.0 & 1.5$\pm$0.4 & 6.040$\pm$0.246 & 0.015$\pm$0.001 & Y$-$ & 1, 3 \\
55003 & 43.9 & 2.2$\pm$0.4 & 7.082$\pm$0.420 & 0.007$\pm$0.001 & Y$-$ & 1, 2 \\
55017 & 20.9 & 1.7$\pm$0.4 & 9.669$\pm$0.236 & 0.003$\pm$0.001 & Y$-$ & 1 \\
55021 & 23.1 & 1.3$\pm$0.4 & 10.768$\pm$0.276 & 0.006$\pm$0.001 & Y$-$ & 1 \\
55026 & 20.2 & 1.6$\pm$0.4 & 10.219$\pm$0.167 & 0.002$\pm$0.001 & Y$+$ & 1, 2 \\
55033 & 24.4 & 2.0$\pm$0.5 & 6.178$\pm$0.176 & 0.023$\pm$0.001 & Y$+$ & 1 \\
55034 & 58.2 & 3.3$\pm$0.5 & 8.639$\pm$0.244 & 0.013$\pm$0.001 & Y$-$ & 1, 2 \\
55035 & 52.2 & 2.7$\pm$0.5 & 9.664$\pm$0.234 & 0.011$\pm$0.001 & Y$-$ & 1 \\
55038 & 53.3 & 2.6$\pm$0.5 & 5.939$\pm$0.181 & 0.012$\pm$0.001 & Y$-$ & 1 \\
55041 & 20.5 & 1.5$\pm$0.5 & 5.560$\pm$0.238 & 0.022$\pm$0.001 & Y$+$ & 1 \\
55043 & 32.6 & 2.2$\pm$0.5 & 7.702$\pm$0.266 & 0.026$\pm$0.001 & Y$+$ & 1 \\
55092 & 23.8 & 1.7$\pm$0.4 & 6.789$\pm$0.200 & 0.018$\pm$0.003 & Y$-$ & new \\
55197 & 20.1 & 1.8$\pm$0.5 & 1.308$\pm$0.352 & 0.037$\pm$0.002 & N & new \\
55339 & 22.0 & 2.0$\pm$0.5 & 7.874$\pm$0.560 & 0.008$\pm$0.001 & Y$-$ & 4, 5 \\
55341 & 27.6 & 2.2$\pm$0.5 & 5.830$\pm$0.308 & 0.010$\pm$0.002 & Y$-$ & 5, 6 \\
55586 & 22.1 & 1.4$\pm$0.4 & 5.219$\pm$0.548 & 0.021$\pm$0.001 & Y$+$ & 7 \\
55596 & 32.4 & 2.4$\pm$0.5 & 3.580$\pm$0.450 & 0.018$\pm$0.001 & Y$-$ & 7 \\
55600 & 30.1 & 2.6$\pm$0.6 & 8.650$\pm$0.850 & 0.018$\pm$0.001 & Y$-$ & 7, 8 \\
55607 & 31.5 & 2.0$\pm$0.5 & 6.220$\pm$1.420 & 0.011$\pm$0.002 & Y$-$ & 7 \\
55642 & 51.7 & 2.1$\pm$0.4 & 2.740$\pm$0.750 & 0.006$\pm$0.001 & N & 7, 9, 10 \\
55643 & 27.6 & 1.9$\pm$0.5 & 2.740$\pm$0.750 & 0.018$\pm$0.001 & N & 7 \\
55890 & 20.1 & 1.3$\pm$0.4 & 2.620$\pm$1.050 & 0.036$\pm$0.002 & N & new \\

\enddata

\tablecomments{ Days (MJD) on which the test statistic $\ge$ 20 ($\gtrsim4.5\sigma$) at the position of Cyg~X-3 from a likelihood analysis of \fermi-LAT observations in 0.1--10\,GeV.}
\tablenotetext{a}{Test statistic on this date from \fermi-LAT.}
\tablenotetext{b}{Photon flux on this date from \fermi-LAT ($10^{-6}$\,\phcms).}
\tablenotetext{c}{Count rate near this date from \rxte-ASM (\cps\ in 3--5\,keV).}
\tablenotetext{d}{Count rate near this date from \swift-BAT (\cps\,\cmsq\ in 15--50\,keV).}
\tablenotetext{e}{Whether the X-ray criteria as defined by \citet{corbel12} were satisfied: i.e., if on the listed date, \cygiii\ had a count rate $\geq$3\,\cps\ in the ASM band, and a count rate that passes the critical value of 0.02\,\cps\ in the BAT band (``Y+'' if the BAT rate has risen above 0.02, and ``Y$-$'' if the BAT rate has dropped below 0.02). The ASM data are degraded after MJD\,55200 \citep[e.g.,][]{gri13}.}
\tablenotetext{f}{Contemporaneous gamma-ray detections by \fermi-LAT or by \agile: (1) \citet{corbel09}; (2) \citet{pia12}; (3) \citet{bul12}; (4) \citet{bul10b}; (5) \citet{corbel10}; (6) \citet{wil11}; (7) \citet{corbel12}; (8) \citet{bul11b}; (9) \citet{corbel11}; (10) \citet{bul11c}.}

\label{tab_cygx3}
\end{deluxetable}

%
\begin{deluxetable}{ l c c c c c c }
\tabletypesize{\scriptsize}
\tablewidth{0pt}
\tablecaption{Candidate Detections of Cyg~X-1 by \fermi-LAT}
\tablehead{
\colhead{MJD} & \colhead{TS \tablenotemark{a}}  & \colhead{LAT \tablenotemark{b}}  & \colhead{ASM \tablenotemark{c}} & \colhead{BAT \tablenotemark{d}} & \colhead{state \tablenotemark{e}} & \colhead{reported \tablenotemark{f}} }

\startdata

54735 & 9.3 & 0.8$\pm$0.4 & 5.969$\pm$0.124 & 0.179$\pm$0.007 & hard & \\
54923 & 13.1 & $\sim$0.02 & 6.863$\pm$0.140 & 0.203$\pm$0.007 & hard & \\
55001 & 9.7 & $\sim$0.3 & 12.021$\pm$0.277 & 0.234$\pm$0.009 & intermediate & \\
55057 & 12.5 & 0.3$\pm$0.2 & 6.082$\pm$0.319 & 0.184$\pm$0.007 & hard & \\
55108 & 9.6 & 0.8$\pm$0.3 & 6.050$\pm$0.181 & 0.157$\pm$0.006 & hard & \\
55118 & 13.1 & 0.9$\pm$0.4 & 6.109$\pm$0.149 & 0.166$\pm$0.006 & hard & 1 \\
55137 & 11.9 & 1.1$\pm$0.4 & 6.397$\pm$0.242 & 0.179$\pm$0.007 & hard & \\
55154 & 10.8 & 0.9$\pm$0.3 & 8.006$\pm$0.438 & 0.225$\pm$0.009 & hard & \\
55160 & 10.3 & 0.8$\pm$0.3 & 7.478$\pm$0.257 & 0.180$\pm$0.007 & intermediate & \\
55223 & 9.6 & $\sim$0.2 & 6.813$\pm$0.239 & 0.284$\pm$0.014 & hard & \\
55277 & 17.1 & 1.4$\pm$0.4 & 8.996$\pm$0.176 & 0.267$\pm$0.010 & hard & 2 \\
55283 & 9.9 & $\sim$0.1 & 6.145$\pm$0.199 & 0.187$\pm$0.007 & hard & \\
55292 & 9.4 & 0.4$\pm$0.3 & 5.823$\pm$0.211 & 0.159$\pm$0.006 & hard & \\
55361 & 9.8 & $\sim$0.1 & 6.925$\pm$0.241 & 0.184$\pm$0.007 & hard & \\
55375 & 10.2 & $\sim$0.2 & 9.675$\pm$0.406 & 0.149$\pm$0.006 & hard & 3 \\
55671 & 15.2 & 1.1$\pm$0.4 & 7.300$\pm$1.810 & 0.075$\pm$0.003 & intermediate & \\
55738 & 12.4 & 0.5$\pm$0.3 & 2.275$\pm$0.491 & 0.109$\pm$0.004 & hard & \\
55809 & 11.1 & 1.3$\pm$0.5 & 3.809$\pm$1.030 & 0.039$\pm$0.002 & soft & \\
55913 & 12.4 & 0.3$\pm$0.2 & --- & 0.150$\pm$0.005 & intermediate & \\
55931 & 10.4 & 1.0$\pm$0.5 & --- & 0.178$\pm$0.006 & hard & \\
55941 & 12.5 & 1.0$\pm$0.3 & --- & 0.146$\pm$0.005 & intermediate & \\

\enddata

\tablecomments{ Days (MJD) on which the test statistic $\ge$ 9 ($\gtrsim3\sigma$) at the position of Cyg~X-1 from a likelihood analysis of \fermi-LAT observations in 0.1--10\,GeV.}
\tablenotetext{a}{Test statistic on this date from \fermi-LAT.}
\tablenotetext{b}{Photon flux on this date from \fermi-LAT ($10^{-6}$\,\phcms).}
\tablenotetext{c}{Count rate near this date from \rxte-ASM (\cps\ in 3--5\,keV).}
\tablenotetext{d}{Count rate near this date from \swift-BAT (\cps\,\cmsq\ in 15--50\,keV).}
\tablenotetext{e}{X-ray state as defined by \citet{gri13}. }
\tablenotetext{f}{Contemporaneous gamma-ray detections by \agile: (1) \cite{sab10a}; (2) \cite{bul10a}; (3) \cite{sab10b}.}

\label{tab_cygx1}
\end{deluxetable}

%
\begin{deluxetable}{ l c c c c }
\tabletypesize{\scriptsize}
\tablewidth{0pt}
\tablecaption{Candidate Detections of GRS~1915+105 by \fermi-LAT}
\tablehead{
\colhead{MJD} & \colhead{TS \tablenotemark{a}}  & \colhead{LAT \tablenotemark{b}}  & \colhead{ASM \tablenotemark{c}} & \colhead{BAT \tablenotemark{d}} }

\startdata

54868 & 9.7 & 0.023$\pm$0.017 & 33.070$\pm$0.300 & 0.034$\pm$ 0.014 \\
55110 & 14.4 & 1.340$\pm$0.487 & 28.393$\pm$0.506 & 0.068$\pm$0.003 \\
55375 & 12.0 & 1.223$\pm$0.520 & 14.726$\pm$0.482 & 0.068$\pm$0.004 \\
55685 & 10.4 & 1.543$\pm$0.572 & 15.965$\pm$0.970 & 0.070$\pm$0.003 \\

\enddata

\tablecomments{ Days (MJD) on which the test statistic $\ge$ 9 ($\gtrsim3\sigma$) at the position of GRS~1915+105 from a likelihood analysis of \fermi-LAT observations in 0.1--10\,GeV.}
\tablenotetext{a}{Test statistic on this date from \fermi-LAT.}
\tablenotetext{b}{Photon flux on this date from \fermi-LAT ($10^{-6}$\,\phcms).}
\tablenotetext{c}{Count rate near this date from \rxte-ASM (\cps\ in 3--5\,keV).}
\tablenotetext{d}{Count rate near this date from \swift-BAT (\cps\,\cmsq\ in 15--50\,keV).}

\label{tab_grs1915}
\end{deluxetable}

%
\begin{deluxetable}{ l c c c c }
\tabletypesize{\scriptsize}
\tablewidth{0pt}
\tablecaption{Candidate Detections of GX~339$-$4 by \fermi-LAT}
\tablehead{
\colhead{MJD} & \colhead{TS \tablenotemark{a}}  & \colhead{LAT \tablenotemark{b}}  & \colhead{ASM \tablenotemark{c}} & \colhead{BAT \tablenotemark{d}} }

\startdata

54712 & 9.0 & 0.576$\pm$0.333 & $-$0.064$\pm$0.230 & 0.003$\pm$0.001 \\
54833 & 9.3 & 1.139$\pm$0.485 & $-$0.204$\pm$0.667 & 0.003$\pm$0.003 \\
55266 & 11.5 & 1.301$\pm$0.507 & 2.076$\pm$0.260 & 0.061$\pm$0.002 \\
55304 & 9.7 & 0.639$\pm$0.291 & 17.390$\pm$2.950 & 0.025$\pm$0.001 \\
55384 & 15.3 & 1.190$\pm$0.399 & 9.665$\pm$0.496 & 0.001$\pm$0.001 \\
55418 & 12.3 & 1.359$\pm$0.498 & 6.283$\pm$0.514 & 0.000$\pm$0.001 \\
55528 & 12.0 & 0.929$\pm$0.426 & 0.470$\pm$1.000 & 0.018$\pm$0.007 \\
55810 & 9.2 & 1.073$\pm$0.445 & 2.440$\pm$1.780 & 0.000$\pm$0.001 \\
56029 & 11.3 & 0.242$\pm$0.203 & --- & 0.001$\pm$0.001 \\
56032 & 10.4 & 1.017$\pm$0.374 & --- & 0.000$\pm$0.001 \\
56059 & 11.2 & 1.291$\pm$0.477 & --- & 0.000$\pm$0.001 \\

\enddata

\tablecomments{ Days (MJD) on which the test statistic $\ge$ 9 ($\gtrsim3\sigma$) at the position of GX~339$-$4 from a likelihood analysis of \fermi-LAT observations in 0.1--10\,GeV.}
\tablenotetext{a}{Test statistic on this date from \fermi-LAT.}
\tablenotetext{b}{Photon flux on this date from \fermi-LAT ($10^{-6}$\,\phcms).}
\tablenotetext{c}{Count rate near this date from \rxte-ASM (\cps\ in 3--5\,keV).}
\tablenotetext{d}{Count rate near this date from \swift-BAT (\cps\,\cmsq\ in 15--50\,keV).}

\label{tab_gx339}
\end{deluxetable}

\end{document}